\DeclareUnicodeCharacter{2500}{}
\documentclass[10pt]{article}
\usepackage{xurl} % for URLs in bibliography
\usepackage[utf8]{inputenc}
\usepackage[T1]{fontenc}
\usepackage{geometry}
\usepackage{lmodern,microtype}
\usepackage{amsmath,amssymb,amsfonts,amsthm,mathtools}
\usepackage{bm,bbm}
\usepackage{graphicx}
\usepackage{booktabs}
\usepackage{enumitem}
\usepackage{algorithm}
\usepackage{algpseudocode}
\usepackage{hyperref}
\usepackage[nameinlink,capitalize]{cleveref}
\usepackage{tikz-cd}
\usepackage{subcaption} % in preamble
\usepackage[font=small,labelfont=bf]{caption}
\usepackage[sort&compress,numbers]{natbib}
\usepackage{xcolor}   % For comments
\usepackage{lineno}   % Line numbers for review
\usepackage{amsmath, amssymb, amsthm, mathrsfs}

\usepackage[utf8]{inputenc}
\usepackage{booktabs}   % For professional looking tables
\usepackage{multirow}   % For merging rows
\usepackage[table]{xcolor} % For coloring rows
\usepackage{array}      % For better column definitions

\definecolor{headerbeige}{RGB}{220, 222, 210}
\definecolor{finetunegreen}{HTML}{B0D7AA} %

\usepackage{amsmath}
\usepackage{amssymb}
\usepackage{algorithm}
\usepackage{algpseudocode}
\usepackage{bm}
\usepackage{booktabs}
\usepackage{caption}
\title{MathAgent: Multi-Agent Optimization of Mathematical Invariants for Molecular Property Prediction}

\author{Yiming Ren$^{1}$, 
Josh V. Vermaas$^{2,3}$, 
and Guo-Wei Wei$^{1,4,5,*}$
 \\[2mm]
       \\    % Author name
 		\\
        $^1$Department of Mathematics,\\
		  University of Georgia, Athens, GA 30602, USA. \\
         $^2$Department of Biochemistry and Molecular Biology, \\ Michigan State University, East Lansing, MI, 48824, USA.\\ 
        $^3$ MSU-DOE Plant Research Laboratory,\\
         Michigan State University, East Lansing, MI, 48824, USA.\\ 
       $^4$Department of Biochemistry and Molecular Biology,\\
		University of Georgia, Athens, GA 30602, USA. \\
         $^5$School of Computing,\\
		University of Georgia, Athens, GA 30602, USA. 
  	} 

\date{}
\begin{document}

\maketitle

\begingroup
\renewcommand{\thefootnote}{\fnsymbol{footnote}}
%\footnotetext[2]{The first two authors contributed equally to this work.}
\footnotetext[1]{Corresponding author. Email: guowei.wei@uga.edu}
\endgroup

\begin{abstract}
Mathematical invariants provide complementary representations of molecular structure, but their predictive utility can vary across datasets and prediction tasks. For a new dataset, appropriate construction of invariant representations from algebraic topology, differential geometry, topological spectral theory, and commutative algebra is  a challenging task, particularly when evaluating multiple representations is computationally expensive. We develop a Mathematical Agentic Model (MathAgent) to automatically optimize invariant representations in molecular property prediction. The framework combines dataset-aware representation design, low-cost candidate screening, and progressive evaluation. Language-model reasoning supports task interpretation, strategy proposal, and workflow coordination, while scientific computations and numerical decisions are carried out by deterministic tools.
We evaluate MathAgent on protein--ligand binding-affinity and quantitative toxicity prediction tasks. Different datasets favor different combinations of mathematical invariants, demonstrating the need for dataset-adaptive representation selection rather than a fixed universal representation. Objective-conditioned executions further show that the framework can adjust its evaluation strategy according to user-specified goals. MathAgent therefore provides a systematic, evidence-driven, and traceable approach for optimizing mathematical invariant  representations and reducing reliance on manual configuration  and/or   experimentation in molecular property prediction.

\end{abstract}

\newpage

\section{Introduction}

Accurate molecular property prediction is central to multiple applications, including drug discovery, protein
engineering, molecular design, and predictive toxicology
\cite{butler2018machine,vamathevan2019applications}. The success of  \emph{in silico}  property predictions depends
critically on how molecules are represented mathematically. Conventional fixed-length molecular representations, including physicochemical
descriptors such as molecular weight, lipophilicity, polar surface area, and
atom counts, as well as molecular fingerprints, provide compact summaries of
molecular structure and composition \cite{todeschini2008handbook, rogers2010extended}. In contrast, graph-based and geometric
models encode relational and three-dimensional spatial information more
explicitly
\cite{bartok2013representing,gilmer2017neural,
satorras2021n,fuchs2020se}.
% Multiple conventional descriptors provide
% compact structural summaries, whereas graph-based and geometric models encode
% relational and spatial information more explicitly
% \cite{bartok2013representing,gilmer2017neural,
% satorras2021n,fuchs2020se}. 
Mathematical invariants offer unique and explainable molecular  representations through algebraic topology \cite{edelsbrunner2002topological,zomorodian2004computing,
cang2018representability,papamarkou2024position,chen2024multiscale}, topological spectral theory \cite{wang2020persistent,qiu2023persistent,wei2025persistent}, differential  geometry \cite{nguyen2019dg}, discrete curvature \cite{wee2021forman}, and commutative algebra
\cite{suwayyid2025persistent,suwayyid2026cakr}. Because these
formalisms encode different aspects of molecular organization, they provide
complementary rather than interchangeable structural information, and existing machine learning methods in the literature utilize many of these information encodings to make their predictions \cite{nguyen2020review,su2025topological, wee2025review}.

However, introducing too many representations may result in overfitting to a given molecular dataset, and so determining which
mathematical invariants should be used remains a nontrivial modeling problem.
Different invariant families characterize distinct structural properties
through different mathematical constructions, and their relative predictive
utility may therefore vary across datasets and prediction tasks \cite{cang2018representability,nguyen2019dg,wee2021forman,
suwayyid2025persistent}. Their implementation may also require
dataset-specific choices of element groups, interaction domains, filtration
ranges, and multiscale parameters
\cite{nguyen2019dg,wee2021forman,suwayyid2025persistent}.  As a
result, representation selection involves not only choosing appropriate
invariant families but also determining how they should be configured for the
data at hand. In practice, these decisions can require substantial expert
judgment, repeated experimentation, and extensive comparison. The associated
computational burden grows with the number of candidate representations,
particularly when multiscale mathematical features must be generated for
multiple alternative configurations.

These challenges motivate adaptive experimentation for the underlying  mathematical
representation used to describe the molecules. Rather than fixing a representation strategy
entirely \emph{a priori}, candidate representations can be progressively prioritized
using dataset characteristics and empirical evidence obtained during the
modeling process. Such a workflow requires stateful coordination of intermediate results and specialized computations, enabling subsequent modeling decisions to adapt to the evidence accumulated during execution.

Large language model (LLM)-based agents provide a suitable computational
paradigm for coordinating this type of sequential scientific workflow.
Agentic models can maintain and update workflow state, invoke specialized
external tools, incorporate intermediate computational results, and use this
information to guide subsequent actions. General reasoning--action, memory,
and agentic frameworks have established these capabilities
\cite{yao2022react,shinn2023reflexion,wu2023autogen}, while related systems
have extended agentic workflows to machine-learning experimentation and
automated scientific research
\cite{huang2023mlagentbench,guo2024ds,lu2024ai}. These developments are
particularly relevant to scientific problems in which high-level reasoning
must be coordinated with domain-specific computation rather than performed
entirely by the language model. This paradigm has also begun to appear in molecular sciences.
ChemCrow, CACTUS, and Co-scientist combine language-model reasoning with
specialized computational or experimental tools
\cite{m2024augmenting,mcnaughton2024cactus,boiko2023autonomous}, whereas
SciToolAgent, SciAgents, and related multi-agent approaches address scientific
tool orchestration, scientific discovery, and molecular analysis and design
\cite{ding2025scitoolagent,ghafarollahi2025sciagents,
stewart2025molecular}. These studies demonstrate the potential
of agentic models to coordinate iterative scientific workflows in which
specialized tools perform domain computations and their outputs inform
subsequent reasoning and actions.

Despite this progress, the studies considered above do not directly address
the dataset-adaptive selection and configuration of mathematical
representations. This problem differs from general scientific tool
orchestration because the relative utility of alternative representations is
not known \emph{a priori} and must be assessed from dataset characteristics and
empirical evidence. Exploring across different strategies to overcome this challenge therefore motivates a specialized agentic framework
for adaptive mathematical representation selection in molecular property predictions.

To address this problem, we introduce MathAgent (Mathematical Agentic Model), an agentic framework for
adaptive mathematical invariant modeling in molecular property prediction.
MathAgent uses dataset characteristics and empirical evidence to configure,
prioritize, and progressively evaluate candidate mathematical representations. LLM-based
reasoning supports task interpretation, strategy proposal, and workflow
coordination, whereas mathematical feature generation, candidate ranking, and
performance evaluation are carried out by deterministic computational tools.
We evaluate MathAgent across protein--ligand binding-affinity and quantitative
toxicity prediction tasks. Across these benchmarks, different datasets favor
different invariant combinations, indicating that no single mathematical
representation is uniformly preferred across molecular prediction problems.
Objective-conditioned executions further show how the workflow can adapt
candidate evaluation and stopping behavior to different user-specified goals.
More broadly, MathAgent recasts mathematical representation selection from a largely expert-driven process involving manual configuration and repeated
experimentation into a systematic, evidence-driven, and traceable
computational workflow tailored to a specific prediction task. By automating routine representation configuration
and screening while retaining deterministic scientific computation, the
framework allows human expertise to remain focused on scientific problem
formulation, interpretation, and higher-level decision-making.

\section{Methods}

\subsection{Overview of the MathAgent Framework}
\label{sec:framework_overview}

Given a molecular dataset $\mathcal{D}$ and a user request $\mathcal{U}$,
MathAgent adaptively configures, prioritizes, and evaluates mathematical
invariant representations for molecular property prediction. In the present study, the molecular dataset $\mathcal{D}$ comprises molecular
structures, their associated target values, and, when available, predefined
training, validation, and test partitions. The user request $\mathcal{U}$ specifies the predictive
objective and may further define constraints on the admissible invariant
families, the evaluation criterion, and the execution strategy. The available
invariant library is
\[
\mathcal{M}
=
\{\mathrm{CA},\mathrm{PH},\mathrm{PL},\mathrm{EIC},\mathrm{FPRC}\},
\]
where CA denotes commutative algebra based on the persistent
Stanley--Reisner framework \cite{suwayyid2025persistent}, PH persistent
homology \cite{zomorodian2004computing}, PL persistent Laplacian
\cite{memoli2022persistent}, EIC element interactive curvature
\cite{nguyen2019dg}, and FPRC Forman persistent Ricci curvature
\cite{wee2021forman}. These invariant families provide mathematically distinct
descriptions of molecular structure and constitute the representation library
considered in the present framework.

If the user restricts the admissible invariant families to
$\mathcal{M}_{U}\subseteq\mathcal{M}$, the corresponding candidate space is
defined as the set of all nonempty subsets,
\[
\mathcal{C}
=
2^{\mathcal{M}_{U}}\setminus\{\emptyset\}.
\]
For each invariant family $m\in\mathcal{M}_{U}$, let $\Theta_m$ denote its
validated dataset-specific representation configuration, and let
\[
\Theta_c=\{\Theta_m:m\in c\}
\]
denote the configuration associated with a candidate combination
$c\in\mathcal{C}$. MathAgent uses dataset characteristics, training-side
predictive evidence, and the user-specified execution objective to select a
candidate $c^{*}\in\mathcal{C}$ and its associated configuration
$\Theta_{c^{*}}$. The workflow returns the resulting molecular-property
predictions $\hat{\mathbf{y}}$ together with an auditable decision trace
$\mathcal{R}$:
\[
(\mathcal{D},\mathcal{U})
\longrightarrow
\left(c^{*},\Theta_{c^{*}},\hat{\mathbf{y}},\mathcal{R}\right).
\]

\begin{figure}[ht]
    \centering
    \includegraphics[width=\textwidth]{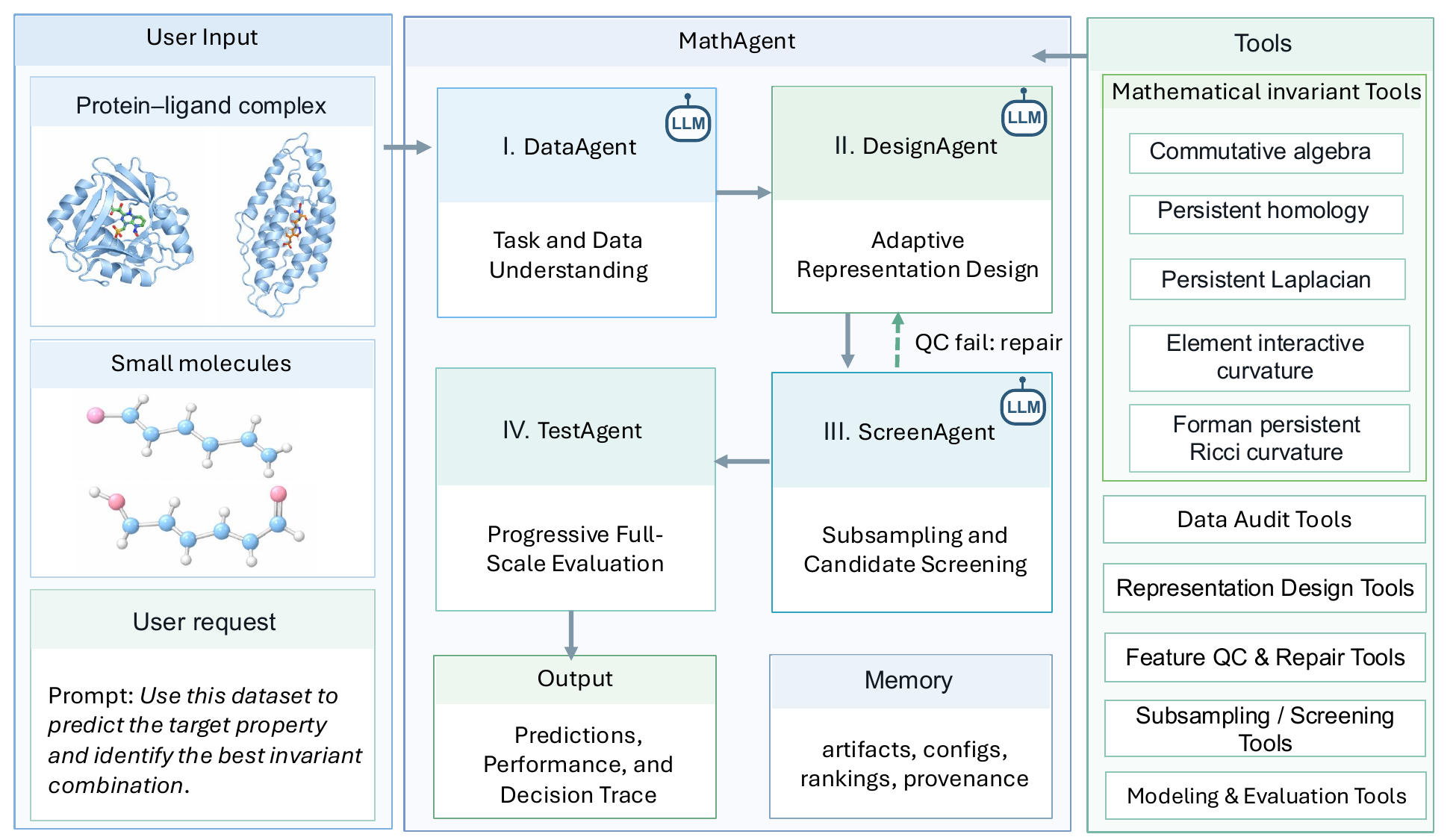}
    \caption{
    Overview of MathAgent for adaptive mathematical invariant modeling in
    molecular property prediction. The framework coordinates task and data
    understanding, dataset-adaptive representation design, training-side candidate
    screening, and progressive full-scale evaluation. LLM-based reasoning
    proposes task-dependent strategies, while deterministic tools validate and
    execute scientific computations. Shared artifact state enables reuse of
    previously generated representations and records the provenance of the
    resulting predictions and representation-selection decisions.
    }
    \label{fig:mitagent_overview}
\end{figure}

As illustrated in Figure~\ref{fig:mitagent_overview}, MathAgent takes as input a
molecular dataset together with a user-specified prediction request, which
defines the task objective and, when provided, constraints on the admissible
representations or execution strategy. MathAgent organizes the workflow into four coordinated functional agents: DataAgent for task and data understanding, DesignAgent for adaptive representation design, ScreenAgent for subsampling and candidate screening, and TestAgent for progressive full-scale evaluation.

DataAgent identifies the prediction objective, available molecular structures,
target values, and data partitions. DesignAgent constructs and validates
dataset-specific configurations for the admissible invariant families.
ScreenAgent evaluates candidate combinations using a representative
training subsample and produces an empirical priority ordering. TestAgent
then promotes high-priority candidates to full-scale evaluation, where the
required invariant representations are generated for the complete
protocol-permitted data partitions rather than only for the training subsample.
Candidates are evaluated according to the screening-derived ordering, while
previously generated compatible representations are reused whenever possible.
The extent of full-scale evaluation is determined by the execution objective:
target-constrained execution may terminate once the prescribed criterion is
satisfied, whereas best-available execution evaluates a predefined candidate
pool before selecting among the evaluated candidates.

Throughout the workflow, LLM-based reasoning supports task interpretation and
strategy proposal, while deterministic tools perform data validation,
representation generation, quality control, candidate ranking, model fitting,
and performance evaluation. These agents are coordinated through a shared
workflow state that stores validated information and reusable computational
artifacts.

\subsection{MathAgent Pipeline}
\label{sec:pipeline}

% To build out the pipeline from Figure~\ref{fig:mitagent_overview}, MathAgent organizes adaptive mathematical
% representation selection into four coordinated stages: Task and Data
% Understanding, Adaptive Representation Design, Subsampling and Candidate Screening, and
% Progressive Full-Scale Evaluation. These stages are coordinated through a shared workflow state that is
% maintained by the orchestration framework.
To implement the pipeline shown in Figure~\ref{fig:mitagent_overview},
MathAgent organizes adaptive mathematical representation selection through four
coordinated functional agents: DataAgent, DesignAgent, ScreenAgent, and TestAgent.
These agents are coordinated through a shared workflow state maintained by the
orchestration framework, which enables validated information and computational
artifacts to be propagated across the workflow.
The state stores validated task information, representation configurations,
screening evidence, and reusable computational artifacts, and can be accessed
and updated by both LLM-based reasoning components and deterministic
computational tools throughout the workflow.

\paragraph{DataAgent: Task and Data Understanding.}
DataAgent converts the user request and the available molecular data into
a well-defined prediction task. The LLM interprets the natural-language request
and proposes a structured task specification that includes the prediction
endpoint, molecular input type (e.g., protein--ligand complexes or small-molecule structures), evaluation metric, admissible invariant
families $\mathcal{M}_U\subseteq\mathcal{M}$, and any user-specified
performance objective or evaluation budget.

The proposed task specification is then checked against the available data by
deterministic data-validation and characterization procedures. These checks verify the presence and
consistency of the required molecular structures, labels, and data partitions,
while also summarizing dataset characteristics relevant to representation
construction, including elemental composition, molecular-size distributions,
interatomic-distance statistics, label distributions, and support for
candidate element and element-pair groups. If essential information is missing or
incompatible with the requested task, the workflow terminates with an explicit
failure state rather than proceeding with an ill-defined prediction problem.

% \paragraph{Adaptive Representation Design.}
\paragraph{DesignAgent: Adaptive Representation Design.}

DesignAgent determines how the admissible mathematical invariants should
be configured for the current dataset. We denote the collection of validated
representation settings by
\[
\Theta=\{\Theta_m:m\in\mathcal{M}_U\},
\]
where $\Theta_m$ specifies the dataset-dependent configuration associated with
invariant family $m$. Depending on the method, these settings may include
element-specific or element-pair groups, filtration ranges, discretization
parameters, and other method-specific quantities.

Using the validated task specification and dataset characteristics from DataAgent,
the LLM proposes dataset-adaptive representation settings for the admissible
invariant families. Deterministic validation procedures examine whether the proposed element groups are sufficiently represented in the dataset, whether parameter values are consistent with the observed structural statistics, and whether method-specific requirements are satisfied before a configuration is accepted for feature
generation. Thus, the LLM provides dataset-aware representation hypotheses,
while the configurations used in numerical computation are determined by
explicit validation rules.

% \paragraph{Subsampling and Candidate Screening.}
\paragraph{ScreenAgent: Subsampling and Candidate Screening.}

ScreenAgent obtains empirical evidence for candidate prioritization before
committing to full-scale representation generation. A training subsample
\[
D_s\subset D_{\mathrm{train}}
\]
is constructed exclusively from the training split of the input dataset(s) and
is designed to preserve task-relevant characteristics of the full training
data, including the target and molecular-size distributions, their joint
variation, and the coverage of supported element and element-pair groups.

The LLM proposes a subsampling strategy based on the task specification
and training-set characteristics, including the candidate subsample size and the
variables used for stratification. The actual subset $D_s$ is then constructed
by deterministic sampling from the resulting strata so that the sampled
proportions approximately follow those of the full training set. The resulting
subsample is checked for training-only membership, agreement with the target and
molecular-size distributions, and structural coverage. Validation,
acceptance, and test samples are excluded from subsample construction.

Representations for the training subsample are generated according to the validated settings
in $\Theta$ and subjected to quality-control and filtration checks before
predictive screening. If a representation fails a repairable validity check,
the affected configuration is returned to DesignAgent for revision, regeneration,
and re-evaluation, while representations that have already passed validation
are retained. This feedback loop is triggered by representation validity or
feature quality rather than predictive performance; a valid representation
that performs poorly remains eligible for empirical comparison.

Only candidate combinations composed entirely of validated representations
proceed to predictive screening. All eligible candidates are evaluated on the
same training subsample using a common downstream modeling protocol and shared
cross-validation partitions. Candidate ordering is determined primarily by predictive performance, with ranking stability and computational cost used as secondary criteria. Resampling-based ranking stability is used to
distinguish closely matched candidates, with computational cost serving as a
subsequent tie-breaking consideration. Numerical ranking is performed by
deterministic tools. The resulting candidate priority order is denoted by
\[
\Pi=(c_1,c_2,\ldots).
\]
This ordering defines the candidate sequence for progressive full-scale
evaluation. For best-available execution, the leading $K$ candidates are fixed
from $\Pi$ before full-scale evaluation begins.

% \paragraph{Progressive Full-Scale Evaluation.}
\paragraph{TestAgent: Progressive Full-Scale Evaluation.}
TestAgent promotes high-priority candidates from candidate screening to
full-scale evaluation while avoiding redundant representation generation.
Before evaluating a candidate, the workflow identifies which constituent
invariant blocks are already available under compatible validated
configurations and computes only the missing components. For example, if PH
and PL have already been generated and the next candidate is PH+PL+CA, only
the CA representation needs to be newly acquired.

The extent of full-scale evaluation is determined by the execution objective.
Under target-constrained execution, the user specifies an evaluation metric and
a corresponding performance target, such as $\mathrm{PCC}^2 \ge 0.60$.
Candidates are evaluated sequentially according to the screening-derived ordering
$\Pi$, and evaluation stops as soon as a candidate satisfies the prescribed
target. If no candidate meets the target after the queue has been exhausted,
the best-performing evaluated candidate under the specified evaluation metric
is returned together with a target-not-reached status. Under best-available
execution, no performance threshold is required; instead, the leading $K$
candidates in $\Pi$ are fixed before full-scale evaluation begins, all $K$
candidates are evaluated, and the best-performing candidate under the specified
evaluation metric is selected.

The execution objective determines how candidates are evaluated and when
evaluation stops, whereas the evaluation protocol determines which data may be
used to guide those decisions. Under the protected-test protocol, the benchmark
test set is excluded from all adaptive representation-selection decisions,
including stopping and candidate comparison. The final representation
combination $c^\star$ is selected using only training- and validation-side
information, after which the protected test set is queried once for final
benchmark reporting.

For objective-conditioned adaptive executions, the benchmark test partition is
instead repurposed as an acceptance set and may be queried during candidate
comparison or stopping decisions. Because this partition participates in
adaptive execution, the resulting scores are interpreted as acceptance-set
performance rather than protected-test benchmark estimates. The acceptance set
may therefore be used to determine whether a prescribed target has been reached
or to compare candidates within a frozen top-$K$ pool.

The workflow returns the selected combination $c^\star$, its validated
representation configuration, molecular-property predictions, and associated
performance records, while the decision trace $\mathcal{R}$ records the
selection history and workflow provenance. Figure~\ref{fig:user_guided_modes}
illustrates the different evaluation trajectories induced by
target-constrained and best-available objectives.

\subsection{MathAgent Components}
\label{sec:components}

The four-agent MathAgent workflow is supported by a set of computational components for
mathematical representation generation, state and artifact management,
quality control, and execution control. These components provide the
infrastructure required to implement the adaptive workflow described in
Section~\ref{sec:pipeline}.

\paragraph{Mathematical invariant tool library.}

The representation layer consists of deterministic generators for the five
mathematical invariant families defined above: persistent homology (PH),
persistent Laplacian (PL), commutative algebra (CA), element interactive
curvature (EIC), and Forman persistent Ricci curvature (FPRC)
\cite{zomorodian2004computing,memoli2022persistent,
suwayyid2025persistent,nguyen2019dg,wee2021forman}. Each tool
accepts a validated dataset-specific configuration $\Theta_m$ and produces a
feature artifact that can be stored, evaluated, and reused across candidate
combinations. A common interface allows these mathematically distinct
invariant families to be invoked within the same workflow while preserving
their method-specific constructions. The mathematical definitions and feature
constructions are given in
Section~\ref{sec:mathematical_invariant_descriptors}.

\paragraph{Shared workflow state and artifact memory.}

\begin{figure*}[!t]
    \centering
    \includegraphics[width=\textwidth]{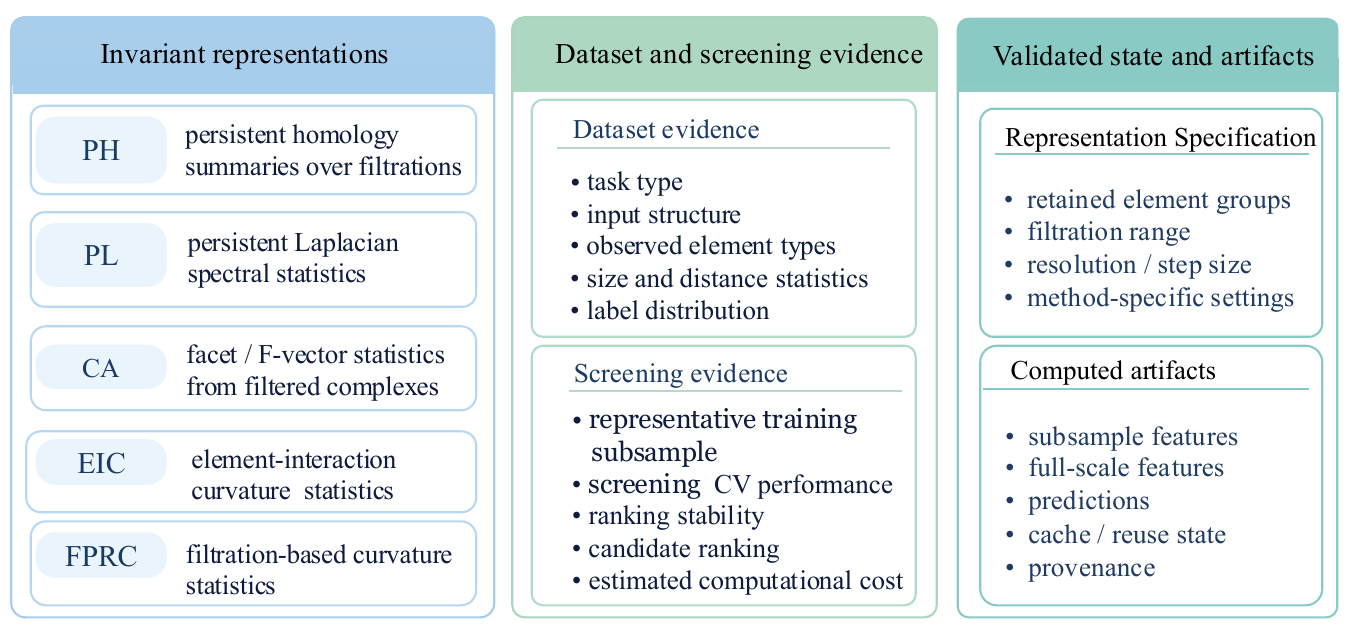}
    \caption{
    Shared state and artifact memory in MathAgent.
    The maintained records include invariant representation, dataset and screening
    evidence, validated configurations, and reusable computational artifacts.
    Provenance metadata supports compatibility checks and reuse across the workflow.
    }
    \label{fig:shared_state}
\end{figure*}

MathAgent maintains a shared state that propagates validated task information,
computational evidence, and reusable artifacts across agents. Following recent
work on persistent and provenance-aware memory in agentic systems
\cite{wang2026map,hsu2026ea}, the maintained information includes invariant
representation metadata, dataset and screening evidence, validated representation
configurations, candidate priorities, generated invariant features, and their
associated provenance. As summarized in Figure~\ref{fig:shared_state}, this
state provides a common record of validated intermediate results throughout the
workflow.

Reusable artifacts include subsample and full-scale feature blocks, predictions,
quality-control records, and evaluation results. Each artifact records the
dataset, representation configuration, and computational settings under which
it was generated. Artifacts are reused only when these conditions are
compatible with the current computation, thereby avoiding redundant generation
of unchanged invariant blocks while preserving traceability.

\paragraph{Representation quality control.}

A deterministic quality-control layer verifies that generated representations
are numerically and structurally suitable for predictive modeling. The checks
include non-finite or missing values, degenerate or constant features, excessive
sparsity, and method-specific filtration consistency. Failed checks trigger the
representation repair mechanism described in Section~\ref{sec:pipeline},
whereas predictive performance is evaluated separately through screening-based
candidate ranking.

\paragraph{Execution control.}
A deterministic execution controller applies the candidate ordering, stopping
rules, and evaluation protocol defined in Section~\ref{sec:pipeline}. It also
coordinates full-scale evaluation with the artifact memory so that compatible
representations are reused and only missing invariant blocks are generated.

Together, the invariant tool library, shared state and artifact memory,
quality-control layer, and execution controller provide the computational
infrastructure supporting the four-agent MathAgent workflow.
Algorithm~\ref{alg:mimagent} summarizes how these components are coordinated
across DataAgent, DesignAgent, ScreenAgent, and TestAgent during execution.

\begin{algorithm}[htp]
\caption{MathAgent execution workflow}
\label{alg:mimagent}
\begin{algorithmic}[1]

\Require Molecular dataset $\mathcal{D}$, user request $\mathcal{U}$,
invariant library $\mathcal{M}$
\Ensure Selected invariant combination $c^\star$,
validated configuration $\Theta_{c^\star}$,
predictions $\hat{\mathbf{y}}$, workflow trace $\mathcal{R}$

% \Statex \textit{Stage I: Task and data understanding}
\Statex \textit{ I: DataAgent -- Task and Data Understanding}
\State $(\mathcal{S}_{\mathrm{task}},\mathcal{S}_{\mathrm{data}})
\gets
\textsc{InterpretAndAudit}
(\mathcal{U},\mathcal{D},\mathcal{M})$

\State Extract admissible methods $\mathcal{M}_U$,
execution mode $e$, target $\tau$, top-$K$ budget $K$,
and evaluation protocol $q$

\If{task or data validation fails}
    \State \Return \textsc{FailureReport}
\EndIf
\State Initialize or restore reusable artifact memory $\mathcal{A}$

\State Define protocol-permitted evaluation data $D_{\mathrm{eval}}$

\If{$q=\texttt{PROTECTED}$}
    \State Exclude $D_{\mathrm{test}}$ from adaptive selection
    and stopping
\EndIf

% \Statex \textit{Stage II: Adaptive representation design}
\Statex \textit{ II: DesignAgent -- Adaptive Representation Design}
\State $\widetilde{\Theta}
\gets
\textsc{LLMProposeRepresentation}
(\mathcal{S}_{\mathrm{task}},
 \mathcal{S}_{\mathrm{data}},
 \mathcal{M}_U)$

\State $\Theta
\gets
\textsc{ValidateRepresentation}
(\widetilde{\Theta},
 \mathcal{S}_{\mathrm{data}})$

% \Statex \textit{Stage III: Subsampling and Candidate Screening}
\Statex \textit{ III: ScreenAgent -- Subsampling and Candidate Screening}
\State $\widetilde{S}
\gets
\textsc{LLMProposeSubsamplingStrategy}
(\mathcal{S}_{\mathrm{task}},
 \mathcal{S}_{\mathrm{data}},
 \Theta)$

\State $D_s
\gets
\textsc{ConstructAndValidateSubsample}
(\widetilde{S},D_{\mathrm{train}})$

\State $(\mathcal{M}_{\mathrm{valid}},
\mathcal{F}_s,\Theta)
\gets
\textsc{GenerateValidateRepair}
(D_s,\mathcal{M}_U,\Theta)$

\State $\mathcal{C}_{\mathrm{valid}}
\gets
2^{\mathcal{M}_{\mathrm{valid}}}\setminus\{\emptyset\}$

\State $\Pi
\gets
\textsc{RankOnSubsample}
(\mathcal{C}_{\mathrm{valid}},\mathcal{F}_s)$

\State $\Pi_{\mathrm{eval}}
\gets
\begin{cases}
\textsc{TopK}(\Pi,K), & e=\texttt{BEST},\\
\Pi, & e=\texttt{TARGET}.
\end{cases}$

% \Statex \textit{Stage IV: Progressive full-scale evaluation}
\Statex \textit{ IV: TestAgent -- Progressive Full-Scale Evaluation}

\ForAll{$c\in\Pi_{\mathrm{eval}}$}

    \State Acquire only invariant blocks in $c$
    that are missing or incompatible in $\mathcal{A}$

    \State $(\hat{\mathbf{y}}^{\,\mathrm{eval}}_c,
    r^{\mathrm{eval}}_c)
    \gets
    \textsc{EvaluateCandidate}
    (c,\mathcal{A},D_{\mathrm{eval}})$

    \State Store evaluation record and update $\mathcal{A}$

    \If{$e=\texttt{TARGET}$ and
    $\textsc{Accept}(r^{\mathrm{eval}}_c,\tau)$}
        \State $c^\star\gets c$
        \State \textbf{break}
    \EndIf

\EndFor

\If{$c^\star$ is undefined}
    \State $c^\star
    \gets
    \textsc{BestEvaluatedCandidate}
    (\{(c,r_c^{\mathrm{eval}}):
    c\in\Pi_{\mathrm{eval}}\})$
\EndIf

\State Freeze $c^\star$ and its configuration $\Theta_{c^\star}$

\If{$q=\texttt{PROTECTED}$}

    \State Acquire any missing test-set artifacts required by $c^\star$

    \State $(\hat{\mathbf{y}},r_{\mathrm{test}})
    \gets
    \textsc{FinalEvaluate}
    (c^\star,\mathcal{A},D_{\mathrm{test}})$
    \Comment{single protected-test evaluation}

\Else

    \State $\hat{\mathbf{y}}
    \gets
    \hat{\mathbf{y}}^{\,\mathrm{eval}}_{c^\star}$

\EndIf

\State $\mathcal{R}
\gets
\textsc{BuildTrace}
(c^\star,\Theta_{c^\star},\Pi,\mathcal{A},e,q)$

\State \Return
$(c^\star,\Theta_{c^\star},\hat{\mathbf{y}},\mathcal{R})$

\end{algorithmic}
\end{algorithm}

\subsection{Mathematical Invariants}
\label{sec:mathematical_invariant_descriptors}

MathAgent permits mathematical constructions from five
invariant families introduced above: commutative algebra (CA), persistent
homology (PH), persistent Laplacian (PL), element interactive curvature (EIC),
and Forman persistent Ricci curvature (FPRC). These representations describe
molecular structure from complementary algebraic, topological, spectral, and
geometric perspectives. Their dataset-specific implementations follow the
validated representation configurations established in
Section~\ref{sec:pipeline}, and the resulting descriptors are organized into
feature blocks for downstream predictive modeling.

Although the five constructions differ mathematically, each is designed to
preserve structural information under transformations appropriate to its
definition, such as label-preserving reindexing, Euclidean rigid motion, or
simplicial isomorphism. The corresponding invariance properties and
scale-dependent constructions are specified separately below. PH characterizes
the persistence of homological structure across a filtration; PL augments
topological information with simplicial spectral structure; CA encodes
algebraic--combinatorial properties of simplicial complexes; EIC describes
continuous geometry through element-resolved interaction-density fields; and
FPRC characterizes local combinatorial curvature across filtration scales.

\subsubsection{Simplicial Complexes and Filtrations}
\label{sec:simplicial_complexes_filtration}

PH, PL, CA, and FPRC share a common construction based on filtered
simplicial complexes. We therefore introduce the required notation before
describing the individual invariant families. EIC is based instead on
continuous interaction-density fields and is treated separately below.

A $q$-simplex is the convex hull of $q+1$ affinely independent vertices.
Thus, simplices of dimensions $0$, $1$, $2$, and $3$ correspond,
respectively, to vertices, edges, filled triangles, and tetrahedra. A
simplicial complex $\mathcal{K}$ is a finite collection of simplices that is
closed under the face relation: whenever $\sigma\in\mathcal{K}$, every face
of $\sigma$ also belongs to $\mathcal{K}$.

Let $\Bbbk$ be a field. The $q$-chain group
$C_q(\mathcal{K};\Bbbk)$ is the vector space generated by the oriented
$q$-simplices of $\mathcal{K}$. The boundary operator is the linear map
\[
\partial_q:
C_q(\mathcal{K};\Bbbk)
\longrightarrow
C_{q-1}(\mathcal{K};\Bbbk),
\]
whose action on an oriented simplex is
\[
\partial_q[u_0,\ldots,u_q]
=
\sum_{\ell=0}^{q}
(-1)^\ell
[u_0,\ldots,\widehat{u_\ell},\ldots,u_q],
\]
where $\widehat{u_\ell}$ indicates that the vertex $u_\ell$ is omitted.
The identity $\partial_q\partial_{q+1}=0$ implies that every boundary is
also a cycle. Consequently, the $q$-th homology group is
\[
H_q(\mathcal{K};\Bbbk)
=
\frac{\ker\partial_q}{\operatorname{im}\partial_{q+1}},
\]
and the corresponding Betti number is
\[
\beta_q
=
\dim_{\Bbbk} H_q(\mathcal{K};\Bbbk).
\]
Geometrically, $\beta_0$, $\beta_1$, and $\beta_2$ count independent
connected components, one-dimensional loops, and enclosed two-dimensional
cavities, respectively.

\begin{figure}[htp]
    \centering
    \includegraphics[width=0.7\linewidth]{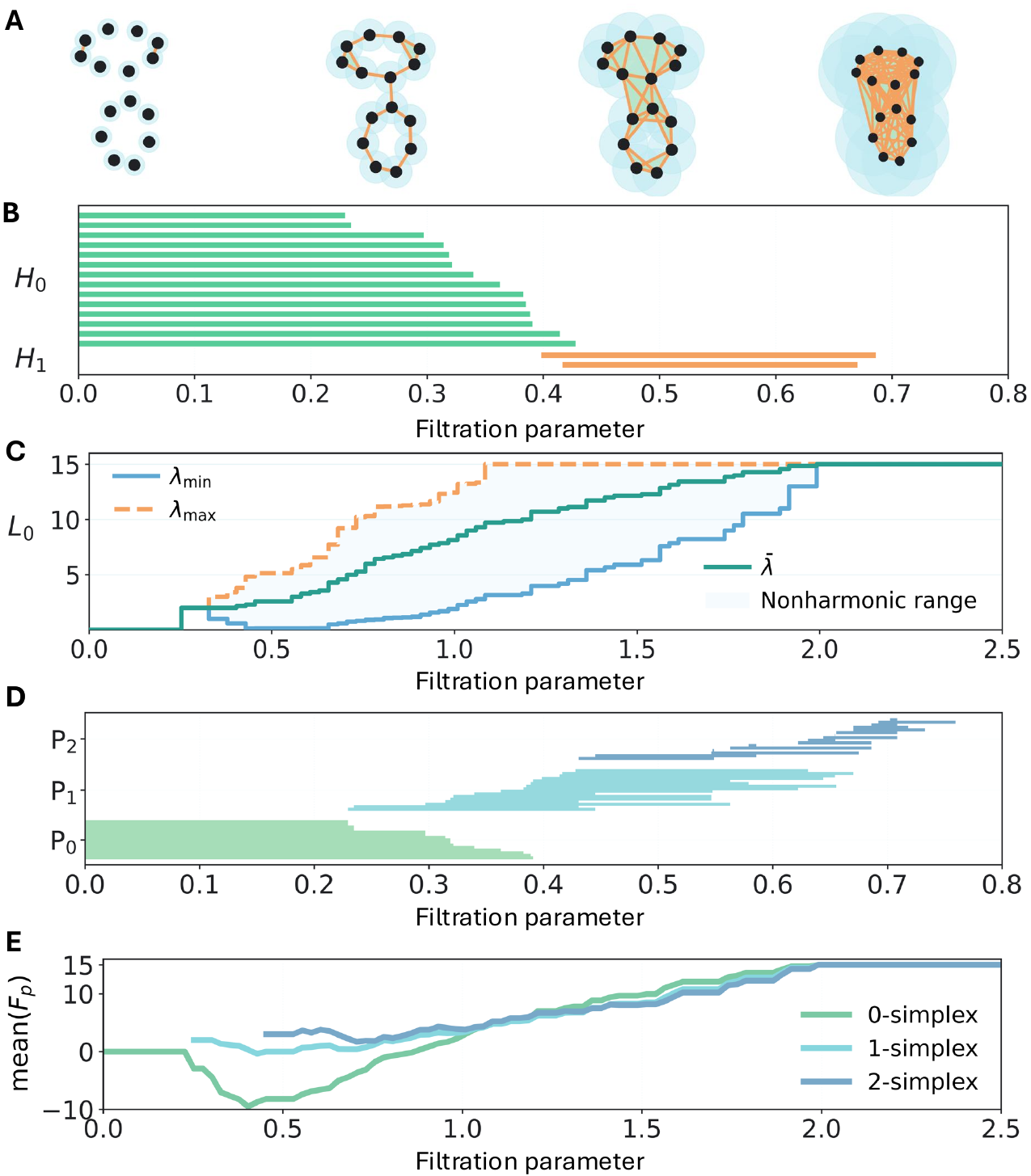}
    \caption{
    Schematic comparison of the four filtration-based mathematical invariant
    families in MathAgent using a common Vietoris--Rips filtration of a
    14-point point cloud containing two loop-like structures.
    (A) Representative simplicial complexes obtained as the filtration
    parameter increases.
    (B) Persistence intervals in dimensions $0$ and $1$, illustrating the
    PH representation.
    (C) Scale-dependent nonharmonic spectral information from the
    $0$-dimensional Laplacian $L_0$, represented by the smallest and largest
    positive eigenvalues, the mean positive eigenvalue
    $\bar{\lambda}$, and the positive spectral range, illustrating PL.
    (D) Persistence bars $P_0$, $P_1$, and $P_2$ associated with
    $0$-, $1$-, and $2$-dimensional facets, illustrating a component of
    the CA representation.
    (E) Evolution of the signed mean Forman curvature
    $\operatorname{mean}(F_p)$ for $p=0,1,2$, illustrating FPRC.
    EIC is omitted from this schematic because its construction is based on
    continuous element-interaction density fields rather than on the
    simplicial filtration shown here.
    }
    \label{fig:math_invariant_example}
\end{figure}

Multiscale structure is introduced through a filtration, that is, a nested
family of simplicial complexes
\[
\mathcal{K}_{t_0}
\subseteq
\mathcal{K}_{t_1}
\subseteq
\cdots
\subseteq
\mathcal{K}_{t_M},
\qquad
t_0<t_1<\cdots<t_M.
\]
Examples relevant to geometric point-cloud data include the
Vietoris--Rips and alpha filtrations
~\cite{vietoris1927hoheren,edelsbrunner2010computational,
edelsbrunner2011alpha}.
Whenever $t_a\leq t_b$, the canonical inclusion
\[
\mathcal{K}_{t_a}
\hookrightarrow
\mathcal{K}_{t_b}
\]
relates the simplicial structures at the two scales. This nested structure
provides the common mathematical basis for the topological, spectral,
algebraic--combinatorial, and discrete-curvature constructions developed
below. Representative stages of such a filtration are illustrated in
Figure~\ref{fig:math_invariant_example}A.

\subsubsection{Persistent Homology (PH)}
\label{sec:persistent_homology}

Persistent homology (PH) \cite{zomorodian2004computing} extends ordinary
simplicial homology by following homological classes as the filtration scale
changes. For indices $a\leq b$, the inclusion
\[
\mathcal{K}_{t_a}
\hookrightarrow
\mathcal{K}_{t_b}
\]
induces the homomorphism
\[
\phi_{a,b}^{q}:
H_q(\mathcal{K}_{t_a};\Bbbk)
\longrightarrow
H_q(\mathcal{K}_{t_b};\Bbbk).
\]
The $q$-dimensional persistent homology group over the interval
$[t_a,t_b]$ is defined by the image of this induced map,
\[
H_q^{a,b}
=
\operatorname{im}\phi_{a,b}^{q},
\]
with persistent Betti number
\[
\beta_q^{a,b}
=
\dim_{\Bbbk}H_q^{a,b}.
\]
Hence, $\beta_q^{a,b}$ measures the number of independent
$q$-dimensional homological classes present at scale $t_a$ that remain
nontrivial at scale $t_b$. Across the complete filtration, the appearance and
disappearance of such classes are represented by persistence intervals,
providing a multiscale description of the underlying topology.

PH is unchanged by vertex relabelings and by simplicial isomorphisms that
preserve the filtration. When the filtration is constructed from pairwise
Euclidean distances, rigid translations and rotations also leave the
persistence barcode unchanged because they preserve all pairwise distances.
Uniform spatial rescaling, in contrast, rescales the filtration parameter and
is therefore not included among the invariant transformations considered
here.

Figure~\ref{fig:math_invariant_example}B shows the $H_0$ and $H_1$
persistence intervals of the schematic point cloud. The $H_0$ intervals
track the merging of initially disconnected components, whereas the $H_1$
intervals record the creation and subsequent filling of the two loop-like
structures.

\subsubsection{Persistent Laplacian (PL)}
\label{sec:persistent_laplacian}

Persistent (combinatorial) Laplacian (PL), introduced by Wang et al.  \cite{wang2020persistent},  enriches persistent homology by retaining
both harmonic and nonharmonic spectral information associated with simplicial
complexes across the filtration \cite{wang2021hermes}. For this construction,
the chain groups are taken over $\mathbb{R}$ and equipped with the inner
product for which the oriented simplices form an orthonormal basis. The
$q$-dimensional combinatorial Laplacian is
\begin{equation}
    L_q
    =
    \partial_{q+1}\partial_{q+1}^{*}
    +
    \partial_q^{*}\partial_q,
    \label{eq:ordinary_laplacian}
\end{equation}
where $\partial_q^{*}$ is the adjoint of $\partial_q$.

The operator $L_q$ is real, symmetric, and positive semidefinite. It yields
the discrete Hodge decomposition
\[
    C_q(\mathcal{K};\mathbb{R})
    =
    \operatorname{im}\partial_{q+1}
    \oplus
    \ker L_q
    \oplus
    \operatorname{im}\partial_q^{*}.
\]
Furthermore,
\[
    \ker L_q
    \cong
    H_q(\mathcal{K};\mathbb{R}),
\]
so that the multiplicity of the zero eigenvalue of $L_q$ is precisely the
$q$-th Betti number \cite{eckmann1944harmonische}. The harmonic part of the
spectrum therefore recovers homological information, whereas the positive
eigenvalues encode additional information about the incidence and
connectivity structure of the simplicial complex.

To extend the Laplacian construction across two filtration scales
$t_a\leq t_b$, define
\[
    C_{q+1}^{a,b}
    =
    \left\{
    c\in C_{q+1}(\mathcal{K}_{t_b};\mathbb{R})
    \ \middle|\
    \partial_{q+1}^{b}c
    \in C_q(\mathcal{K}_{t_a};\mathbb{R})
    \right\}.
\]
Restricting the boundary map at $t_b$ to this subspace gives the persistent
boundary operator
\[
    \partial_{q+1}^{a,b}:
    C_{q+1}^{a,b}
    \rightarrow
    C_q(\mathcal{K}_{t_a};\mathbb{R}).
\]
The corresponding persistent Laplacian is
\begin{equation}
    L_q^{a,b}
    =
    \partial_{q+1}^{a,b}
    \bigl(\partial_{q+1}^{a,b}\bigr)^{*}
    +
    \bigl(\partial_q^{a}\bigr)^{*}\partial_q^{a}.
    \label{eq:persistent_laplacian}
\end{equation}
Its null space satisfies
\[
    \ker L_q^{a,b}\cong H_q^{a,b},
\]
and therefore the multiplicity of its zero eigenvalue equals the persistent
Betti number $\beta_q^{a,b}$. In contrast, the positive portion of the
spectrum captures nonharmonic structural information that cannot, in general,
be recovered from persistent Betti numbers alone
~\cite{wang2021hermes,memoli2022persistent,jones2025petls}.

The PL spectrum is invariant under filtration-preserving simplicial
relabelings. Reordering the simplices, or changing their orientations
consistently, modifies the matrix representation of the Laplacian through an
orthogonal signed-permutation similarity transformation,
\[
    L_q'=P_q^{\top}L_qP_q,
\]
which leaves its eigenvalues unchanged. For filtrations determined by
Euclidean distances, rigid translations and rotations likewise preserve the
filtered complexes and therefore their Laplacian spectra.

Figure~\ref{fig:math_invariant_example}C illustrates the evolution of the
positive spectrum of $L_0$. At the initial scale, when all vertices are
isolated, the spectrum contains only harmonic modes. Positive eigenvalues
appear as edges are added and subsequently evolve with the connectivity of
the complex. For the 14-point example, the positive eigenvalues eventually
coincide once the Vietoris--Rips filtration reaches the complete clique
complex.

\subsubsection{Commutative Algebra (CA)}
\label{sec:persistent_commutative_algebra}

The commutative-algebra (CA) representation used in MathAgent is based on
the persistent Stanley--Reisner theory introduced by Suwayyid and Wei
\cite{suwayyid2025persistent}. This construction translates a filtered
simplicial complex into a family of commutative-algebraic objects, thereby
encoding its combinatorial organization through squarefree monomial ideals,
Stanley--Reisner quotient rings, and facet-associated prime ideals.

Let $\Delta$ be a finite simplicial complex with vertex set
$W=\{y_1,\ldots,y_n\}$, and let
$\varphi:\Delta\rightarrow\mathbb{R}$ be a monotone face function satisfying
$\varphi(\omega)\leq\varphi(\omega')$ whenever
$\omega\subseteq\omega'$. The sublevel sets of $\varphi$ define the
filtration
\[
\Delta_\epsilon
=
\{\omega\in\Delta\mid\varphi(\omega)\leq\epsilon\},
\]
for which
$\Delta_{\epsilon_a}\subseteq\Delta_{\epsilon_b}$ whenever
$\epsilon_a\leq\epsilon_b$.

Fix a field $\Bbbk$ and consider the standard graded polynomial ring
$R=\Bbbk[y_1,\ldots,y_n]$. At filtration scale $\epsilon$, the
Stanley--Reisner ideal associated with $\Delta_\epsilon$ is
\begin{equation}
    I_\epsilon
    =
    I(\Delta_\epsilon)
    =
    \left(
    y_{i_1}\cdots y_{i_s}
    \ \middle|\
    \{y_{i_1},\ldots,y_{i_s}\}\notin\Delta_\epsilon
    \right)
    \subseteq R,
    \label{eq:sr_ideal}
\end{equation}
and the corresponding Stanley--Reisner ring is
$\Bbbk[\Delta_\epsilon]=R/I_\epsilon$. Because increasing $\epsilon$
introduces additional simplices into the complex, the collection of nonfaces
decreases, producing a descending family of squarefree monomial ideals.

The facets of $\Delta_\epsilon$ are encoded through the irredundant
minimal-prime decomposition
\begin{equation}
    I_\epsilon
    =
    \bigcap_{\omega\in\mathcal{F}(\Delta_\epsilon)}
    P_\omega,
    \qquad
    P_\omega
    =
    (y_i\mid y_i\notin\omega),
    \label{eq:facet_prime_decomposition}
\end{equation}
where $\mathcal{F}(\Delta_\epsilon)$ denotes the facet set of
$\Delta_\epsilon$. To separate facets according to dimension, define
\[
    \mathcal{P}_\epsilon^{(p)}
    =
    \left\{
    P_\omega
    \ \middle|\
    \omega\in\mathcal{F}(\Delta_\epsilon),
    \ \dim(\omega)=p
    \right\}.
\]
For two scales $\epsilon_a\leq\epsilon_b$, the prime ideals associated
with $p$-dimensional facets that occur at both scales form
\begin{equation}
    \mathcal{P}_{\epsilon_a,\epsilon_b}^{(p)}
    =
    \mathcal{P}_{\epsilon_a}^{(p)}
    \cap
    \mathcal{P}_{\epsilon_b}^{(p)},
    \qquad
    \mu_p^{\epsilon_a,\epsilon_b}
    =
    \left|
    \mathcal{P}_{\epsilon_a,\epsilon_b}^{(p)}
    \right|.
    \label{eq:facet_persistence_number}
\end{equation}
The quantity $\mu_p^{\epsilon_a,\epsilon_b}$ therefore summarizes the
persistence of $p$-dimensional maximal simplices across the two filtration
levels.

Facet persistence characterizes only maximal simplices. To retain information
about simplices of every dimension, let $\Delta_\epsilon^{(p)}$ denote the
set of $p$-simplices in $\Delta_\epsilon$. The associated $f$-vector is
\begin{equation}
    \mathbf{f}^{\epsilon}
    =
    \left(
    f_0^\epsilon,
    f_1^\epsilon,
    \ldots,
    f_{\dim(\Delta_\epsilon)}^\epsilon
    \right),
    \qquad
    f_p^\epsilon
    =
    \left|
    \Delta_\epsilon^{(p)}
    \right|.
    \label{eq:classical_f_vector}
\end{equation}
Facet-persistence quantities and $f$-vectors thus provide complementary
descriptions: the former track maximal simplices, whereas the latter quantify
the full simplex population dimension by dimension.

The CA quantities are invariant under vertex relabeling. A permutation of
the vertices induces the corresponding permutation of the polynomial
variables, hence a graded automorphism of $R$ that carries the associated
Stanley--Reisner ideals to isomorphic ideals. As a result, the
facet-associated quantities and the components of the $f$-vector are
unchanged. These quantities are therefore algebraic--combinatorial
invariants, although they are not complete invariants, since distinct
nonisomorphic simplicial complexes may possess identical facet counts or
$f$-vectors.

Figure~\ref{fig:math_invariant_example}D displays the persistence of
$0$-, $1$-, and $2$-dimensional facets along the filtration. A simplex
enters the facet representation when it becomes maximal and ceases to appear
as a facet once it is contained in a higher-dimensional maximal simplex.
Accordingly, the $P_0$, $P_1$, and $P_2$ bars describe the multiscale
behavior of maximal vertices, edges, and triangles.

\subsubsection{Element Interactive Curvature (EIC)}
\label{sec:element_interactive_curvature}

Element interactive curvature (EIC), introduced by Nguyen and Wei \cite{nguyen2019dg}, represents molecular interactions through
the differential geometry of continuous, element-resolved density fields. Consider a molecular system containing atoms at
positions $\mathbf{x}_j$ with associated weights $w_j$. A continuous field
is obtained from the discrete atomic configuration through
\begin{equation}
    \rho(\mathbf{x};\{\eta_j\},\{w_j\})
    =
    \sum_{j=1}^{N}
    w_j
    \Phi(\|\mathbf{x}-\mathbf{x}_j\|;\eta_j),
    \label{eq:eic_density_reorganized}
\end{equation}
where $w_j=1$ produces a number-density field, whereas $w_j=q_j$ produces
a charge-density field. The correlation kernel $\Phi$ is assumed to be a
monotonically decreasing $C^2$ radial function satisfying
\[
    \Phi(d;\eta)\rightarrow1
    \quad\text{as }d\rightarrow0,
    \qquad
    \Phi(d;\eta)\rightarrow0
    \quad\text{as }d\rightarrow\infty.
\]
Two commonly used choices are the generalized exponential and generalized
Lorentz kernels,
\begin{equation}
\begin{aligned}
    \Phi_{\mathrm{E}}(d;\eta)
    &=
    \exp\left[-\left(\frac{d}{\eta}\right)^{\kappa}\right],
    && \kappa>0, \\
    \Phi_{\mathrm{L}}(d;\eta)
    &=
    \frac{1}{1+\left(\dfrac{d}{\eta}\right)^{\nu}},
    && \nu>0 .
\end{aligned}
\label{eq:eic_correlation_kernels}
\end{equation}

Chemical specificity is introduced by resolving the density field according
to element pairs. For element type $a$, define its van der Waals domain by
\[
D_a
=
\bigcup_{i:e_i=a} B(\mathbf{x}_i,r_i).
\]
For distinct element types $a$ and $b$, the density generated by atoms of
type $b$ over the domain $D_a$ is
\begin{equation}
    \rho_{ab}(\mathbf{x};\eta_{ab})
    =
    \sum_{\substack{j:e_j=b\\
    \|\mathbf{x}_i-\mathbf{x}_j\|>
    r_i+r_j+\sigma,\ \forall\, i:e_i=a}}
    w_j
    \Phi(\|\mathbf{x}-\mathbf{x}_j\|;\eta_{ab}),
    \qquad
    \mathbf{x}\in D_a,
    \label{eq:eic_element_density}
\end{equation}
where $r_i$ and $r_j$ are atomic radii and $\sigma$ denotes the
dataset-dependent correction used in the original EIC formulation. The
distance condition removes covalent contributions from the interaction
density. For interactions involving identical element types, atom-specific
van der Waals domains are used to exclude self-contributions.

Spatial scale is controlled through the characteristic interaction length.
Following the EIC parametrization,
\[
    \eta_{ab}(\tau)
    =
    \tau\bigl(\bar r_a+\bar r_b\bigr),
\]
where $\bar r_a$ and $\bar r_b$ are the van der Waals radii associated with
the two element types and $\tau$ is dimensionless. Varying $\tau$ therefore
changes the spatial extent over which the corresponding element-pair
interaction is represented.

For a fixed pair $(a,b)$ and interaction scale $\eta_{ab}$, an
element-interactive manifold is obtained as a level set of the density field,
\[
    \rho_{ab}(\mathbf{x};\eta_{ab})
    =
    c\,\rho_{\max},
    \qquad
    0\leq c\leq1,
\]
where
\[
\rho_{\max}
=
\max_{\mathbf{x}\in D_a}
\rho_{ab}(\mathbf{x};\eta_{ab}).
\]
At a regular point satisfying $\nabla\rho\neq0$, define
\[
g=\rho_x^2+\rho_y^2+\rho_z^2.
\]
The Gaussian curvature $K$ of the level surface is
\begin{align}
    K
    =&
    \frac{1}{g^2}
    \Big[
    2\rho_x\rho_y\rho_{xz}\rho_{yz}
    +2\rho_x\rho_z\rho_{xy}\rho_{yz}
    +2\rho_y\rho_z\rho_{xy}\rho_{xz}
    \nonumber\\
    &\quad
    -2\rho_x\rho_z\rho_{xz}\rho_{yy}
    -2\rho_y\rho_z\rho_{xx}\rho_{yz}
    -2\rho_x\rho_y\rho_{xy}\rho_{zz}
    \nonumber\\
    &\quad
    +\rho_z^2\rho_{xx}\rho_{yy}
    +\rho_x^2\rho_{yy}\rho_{zz}
    +\rho_y^2\rho_{xx}\rho_{zz}
    -\rho_x^2\rho_{yz}^2
    -\rho_y^2\rho_{xz}^2
    -\rho_z^2\rho_{xy}^2
    \Big],
    \label{eq:eic_gaussian_curvature}
\end{align}
while the mean curvature $H$ is
\begin{align}
    H
    =&
    \frac{1}{2g^{3/2}}
    \Big[
    2\rho_x\rho_y\rho_{xy}
    +2\rho_x\rho_z\rho_{xz}
    +2\rho_y\rho_z\rho_{yz}
    \nonumber\\
    &\quad
    -(\rho_y^2+\rho_z^2)\rho_{xx}
    -(\rho_x^2+\rho_z^2)\rho_{yy}
    -(\rho_x^2+\rho_y^2)\rho_{zz}
    \Big].
    \label{eq:eic_mean_curvature_reorganized}
\end{align}
The two principal curvatures are then
\[
    \kappa_{\min}
    =
    H-\sqrt{H^2-K},
    \qquad
    \kappa_{\max}
    =
    H+\sqrt{H^2-K}.
\]
Together, these curvature quantities describe the local differential geometry
of the element-specific interaction manifold at a prescribed spatial scale.

Because the density fields are constructed from radial functions of Euclidean
distance, relabeling atoms does not alter the resulting fields, provided the
associated element labels and weights are preserved. Rigid translations and
rotations carry the density fields and their level surfaces rigidly through
space without changing their intrinsic curvature values. Gaussian curvature
is independent of the orientation assigned to a level surface, whereas the
sign of mean curvature depends on the chosen normal orientation. Under a
consistent normal convention, the EIC representation is therefore invariant
under atom reindexing and Euclidean rigid motions.

\subsubsection{Forman Persistent Ricci Curvature (FPRC)}
\label{sec:forman_persistent_ricci_curvature}

Forman persistent Ricci curvature (FPRC) describes how discrete
combinatorial curvature changes as a simplicial filtration evolves
~\cite{wee2021forman}. Two distinct $p$-simplices $\alpha$ and
$\bar{\alpha}$ are said to be parallel,
$\bar{\alpha}\parallel\alpha$, when they share a common
$(p-1)$-dimensional face but are not simultaneously faces of any common
$(p+1)$-simplex.

For $p>0$, the Forman Ricci curvature assigned to a $p$-simplex $\alpha$
in an unweighted simplicial complex is
\begin{equation}
    F_p(\alpha)
    =
    \#\{\beta^{p+1}\mid \alpha<\beta^{p+1}\}
    +
    \#\{\gamma^{p-1}\mid \gamma^{p-1}<\alpha\}
    -
    \#\{\bar{\alpha}\mid \bar{\alpha}\parallel\alpha\},
    \label{eq:fprc_frc_reorganized}
\end{equation}
where the first term counts the $(p+1)$-dimensional cofaces containing
$\alpha$, the second counts its $(p-1)$-dimensional faces, and the final
term counts parallel $p$-simplices. The curvature therefore combines local
face--coface incidence with same-dimensional parallelism.

Following the convention adopted in the FPRC construction
~\cite{wee2021forman}, the curvature of a nonisolated vertex $v$ is derived
from the Forman curvatures of the incident edges,
\begin{equation}
    F_0(v)
    =
    \frac{1}{d(v)}
    \sum_{e>v} F_1(e),
    \label{eq:fprc_vertex}
\end{equation}
where $d(v)$ denotes the number of edges incident to $v$. For an isolated
vertex, $F_0(v)$ is set to zero. Evaluating these curvature quantities over
multiple simplex dimensions and filtration values provides a multiscale
description of the changing local combinatorial geometry.

Forman Ricci curvature is determined entirely by local face--coface incidence
and parallel-simplex relations. Because these combinatorial relations are
preserved by simplicial isomorphisms and vertex relabelings, corresponding
simplices retain the same curvature values. When the filtration is generated
from Euclidean distances, translations and rotations additionally preserve
the filtered simplicial complexes. The resulting FPRC quantities are
therefore invariant under these transformations.

Figure~\ref{fig:math_invariant_example}E shows the scale-dependent signed
mean Forman curvature $\operatorname{mean}(F_p)$ for $p=0,1,2$. As the
filtration proceeds, newly introduced simplices modify the local
face--coface and parallel-simplex configurations, producing corresponding
changes in the mean-curvature curves. In the illustrated 14-point example,
the curves eventually coincide and remain constant once the
Vietoris--Rips complex becomes the complete clique complex.

\subsection{Implementation and Experimental Settings}
\label{sec:implementation}

MathAgent was implemented as a LangGraph-based agentic workflow with a shared
state that coordinates language-model reasoning and deterministic scientific
computation. The reported implementation used LangGraph 1.2.11 and GPT-5.6
Terra as the LLM backend. The LLM was restricted to task interpretation,
representation and subsampling strategy proposal, and scientific advisory. Its
outputs were schema constrained and were validated by deterministic Python
procedures before execution; numerical candidate ranking and performance
evaluation were performed independently by deterministic computational tools.

All predictive experiments used a fixed gradient-boosted tree (GBT) learner,
with no hyperparameter optimization performed during representation selection.
The GBT model used 4000 estimators, a learning rate of 0.01, a maximum tree
depth of 7, and a subsampling rate of 0.5. Predictions were averaged across
three runs with random states 42, 43, and 44. Candidate prioritization on the
training subsample was based on shared five-fold out-of-fold predictions and
a deterministic ranking procedure that considered predictive performance,
bootstrap-based ranking stability, computational cost, and fixed tie-breaking
rules. To control stochastic variation, the random seeds for training-set subsampling,
cross-validation, and bootstrap procedures were fixed at 2026.

Representation generation was subject to deterministic quality-control and
filtration checks before candidate evaluation. Invalid or inconsistent feature
artifacts were rejected, while repairable filtration issues were handled by a
deterministic repair procedure before model ranking. For benchmark evaluations
using a protected test set, representation selection was completed using
training-side and, where applicable, validation evidence before the final
representation was fixed and evaluated on the protected test set.

\section{Results}
\label{sec:results}

\subsection{Benchmark Design and Evaluation Protocol}
\label{sec:benchmark_design}

MathAgent was evaluated on nine molecular prediction benchmarks spanning
protein--ligand binding-affinity prediction and quantitative 
toxicity prediction. Five mathematical invariant families were available to
the workflow: persistent homology (PH), persistent Laplacian (PL),
commutative algebra (CA), element interactive curvature (EIC), and Forman
persistent Ricci curvature (FPRC). A common gradient-boosted tree (GBT)-based
downstream modeling protocol was used throughout the present experiments.
This controlled setting allows differences among MathAgent runs to be
examined primarily in terms of adaptive representation construction and
invariant selection rather than changes in predictive architecture.

For protein--ligand binding-affinity prediction, we considered two
complementary benchmark settings. The CASF-2007, CASF-2013, and CASF-2016
benchmarks evaluate scoring power using the PDBbind refined-set/core-set
protocol
\cite{cheng2009comparative,li2014comparative,su2018comparative,
liu2015pdb}. CASF-2007 uses 1,105 training complexes and a 195-complex
core test set, whereas the commonly used CASF-2013 protocol contains
2,764 training complexes and 195 test complexes. CASF-2016 contains 3,772 training complexes and 285 core-set test
complexes \cite{su2018comparative,liu2015pdb}.
We additionally evaluated MathAgent on the LBA30 and LBA60 benchmarks,
which are derived from PDBbind using protein sequence-identity-based
partitioning \cite{townshend2020atom3d}. The reference LBA30 split comprises
3,507 training, 466 validation, and 490 test complexes, whereas the reference
LBA60 split comprises 3,678 training, 460 validation, and 460 test complexes.

For quantitative toxicity prediction, four established endpoints were
considered: IGC50, LC50, LC50DM, and LD50
\cite{wu2018quantitative}. IGC50 measures the concentration required to
inhibit 50\% of the growth of Tetrahymena pyriformis after 40 h.
LC50 denotes the concentration of a test chemical in water that causes 50\%
mortality in fathead minnows after 96 h, whereas LC50DM measures the
corresponding 50\% lethal concentration for Daphnia magna after 48 h.
LD50 quantifies the orally administered dose that results in 50\%
mortality in rats.

Let $y_i$ and $\hat{y}_i$ denote the reference and predicted target values,
respectively, for sample $i$, and let $\bar{y}$ and $\bar{\hat{y}}$ denote
their sample means. The Pearson correlation coefficient is defined as
\[
\mathrm{PCC}(\mathbf{y},\hat{\mathbf{y}})
=
\frac{
\sum_{i=1}^{N}
(y_i-\bar{y})(\hat{y}_i-\bar{\hat{y}})
}{
\sqrt{\sum_{i=1}^{N}(y_i-\bar{y})^2}
\sqrt{\sum_{i=1}^{N}(\hat{y}_i-\bar{\hat{y}})^2}
},
\]
and the root mean squared error is
\[
\mathrm{RMSE}
=
\sqrt{
\frac{1}{N}
\sum_{i=1}^{N}
(y_i-\hat{y}_i)^2
}.
\]
For the toxicity benchmarks, the squared Pearson correlation coefficient is
reported as
\[
\mathrm{PCC}^{2}
=
\left[
\mathrm{PCC}(\mathbf{y},\hat{\mathbf{y}})
\right]^2,
\]
together with RMSE, following the reporting convention adopted in previous
quantitative toxicity studies
\cite{martin2016user,wu2018quantitative,gao20202d}. A consistent evaluation protocol was used across all benchmarks. Candidate representations were selected using training-side evidence, and the selected representation was fixed before the held-out test set was accessed. The reported MathAgent result was then obtained from a single evaluation on the corresponding test set.

\subsection{Protein--Ligand Binding-Affinity Prediction}
\label{sec:binding_results}

\subsubsection{CASF Scoring-Power Benchmarks}
\label{sec:casf_results}

Protein--ligand binding affinity is a key metric in drug discovery, quantifies the interaction strength
between a ligand and its macromolecular target, and is a central quantity in
structure-based drug discovery
\cite{cai2024carsidock,ain2015machine}. Accurate computational prediction of
binding affinity has therefore become an important objective in molecular
modeling and machine learning, and is the first key metric we predict by MathAgent.

While multiple ligand affinity benchmarks are available, our initial focuses for evaluating
MathAgent were the CASF-2007, CASF-2013, and CASF-2016
scoring-power datasets
\cite{cheng2009comparative,li2014comparative,su2018comparative}, so that we can better compare to prior approaches. For each
benchmark, invariant selection was performed exclusively from training-side
evidence, after which the selected representation was fixed and evaluated once
on the corresponding core test set. This protocol preserves a clear separation
between representation selection and final benchmark evaluation. Table~\ref{tab:casf_results_mimagent} compares MathAgent with representative
machine-learning and conventional scoring methods
reported under the corresponding CASF benchmark protocols. Depending on the input dataset, MathAgent selected
different features to build a predictive model, using EIC+FPRC+PH+PL for CASF-2007, EIC+FPRC for CASF-2013, and
EIC+FPRC+PH+PL for CASF-2016.
This variation
shows that the framework does not impose a fixed mathematical representation
across related protein--ligand scoring tasks, but instead adapts the selected
invariant combination to the characteristics of each benchmark dataset.
These different approaches yielded PCC values of 0.825, 0.780, and
0.831 and RMSE values of 1.409, 1.496, and 1.297, respectively (Table~\ref{tab:casf_results_mimagent}).

Across the three CASF benchmarks, MathAgent achieves higher PCC values than
the reference methods included in Table~\ref{tab:casf_results_mimagent}. On
CASF-2007, MathAgent attains a PCC of 0.825, exceeding the reported values for
ChemScore@SYBYL, PLP1@DS, G-Score@SYBYL, PLP2@DS, ASP@GOLD,
ChemScore@GOLD, D-Score@SYBYL, LUDI3@DS, LigScore2@DS, and LUDI2@DS.
On CASF-2013, MathAgent achieves a PCC of 0.780, exceeding PLEC
\cite{wojcikowski2019development}, Pafnucy
\cite{stepniewska2018development}, and the conventional scoring functions
listed in the table, including X-Score, ChemPLP@GOLD, PLP1@DS,
ASP@GOLD, and ChemScore@SYBYL. On CASF-2016, MathAgent achieves a PCC of
0.831, exceeding PLEC, Pafnucy, X-Score, ChemPLP@GOLD, ASP@GOLD,
ChemScore@SYBYL, and the other conventional scoring functions included in the
comparison.

\begin{table*}[htbp]
\centering
\caption{
Comparison of scoring power on the CASF-2007, CASF-2013, and CASF-2016
protein--ligand binding-affinity benchmarks, evaluated by the PCC. Published values are compiled from the corresponding CASF studies
and cited method references
\cite{stepniewska2018development,li2014comparative,su2018comparative,
cheng2009comparative,wang2017improving,
wojcikowski2019development}. A dash indicates that a result was not reported for the
corresponding benchmark.
}
\label{tab:casf_results_mimagent}
\small
\setlength{\tabcolsep}{9pt}
\renewcommand{\arraystretch}{1.08}
\begin{tabular}{@{}lccc@{}}
\toprule
Method & CASF-2007 & CASF-2013 & CASF-2016 \\
\midrule

PLEC \cite{wojcikowski2019development}
    & -- & 0.757 & 0.817 \\

Pafnucy \cite{stepniewska2018development}
    & -- & 0.700 & 0.780 \\

X-Score
    & -- & 0.622 & 0.631 \\

X-ScoreHS
    & -- & 0.620 & 0.629 \\

X-ScoreHM
    & -- & 0.614 & 0.609 \\

X-ScoreHP
    & -- & 0.607 & 0.621 \\

ChemScore@SYBYL
    & 0.555 & 0.592 & 0.590 \\

ChemPLP@GOLD
    & -- & 0.579 & 0.614 \\

PLP1@DS
    & 0.545 & 0.568 & 0.581 \\

G-Score@SYBYL
    & 0.492 & 0.558 & 0.572 \\

PLP2@DS
    & 0.535 & 0.558 & 0.563 \\

ASP@GOLD
    & 0.534 & 0.556 & 0.617 \\

ASE@MOE
    & -- & 0.544 & 0.591 \\

ChemScore@GOLD
    & 0.441 & 0.536 & 0.574 \\

D-Score@SYBYL
    & 0.392 & 0.526 & 0.531 \\

Alpha-HB@MOE
    & -- & 0.511 & 0.569 \\

LUDI3@DS
    & 0.487 & 0.487 & 0.502 \\

Affinity-dG@MOE
    & -- & 0.482 & 0.552 \\

LigScore2@DS
    & 0.464 & 0.456 & 0.540 \\

GlideScore-SP
    & -- & 0.452 & 0.513 \\

LUDI2@DS
    & 0.410 & 0.451 & 0.526 \\

\midrule
MathAgent
    & 0.825 & 0.780 & 0.831 \\

\bottomrule
\end{tabular}
\end{table*}

\subsubsection{Sequence-Generalization on LBA Benchmarks}
\label{sec:lba_results}

The CASF benchmarks assess scoring power under the conventional PDBbind
refined-set/core-set protocol but do not explicitly control protein sequence
similarity across data partitions. To examine generalization to less homologous
protein targets, MathAgent was further evaluated on the Atom3D-derived LBA30
and LBA60 benchmarks, which employ sequence-identity-based partitioning
\cite{townshend2020atom3d}.

For LBA30 and LBA60, MathAgent selected EIC+FPRC and EIC, respectively.
The corresponding PCC values were 0.623 and 0.787, with RMSE values of
1.326 and 1.262. In both cases, representation selection was completed using
training and validation information before access to the held-out test set,
and the fixed representation was subsequently evaluated once on the
corresponding test partition.

On LBA30, MathAgent achieves a PCC of 0.623 and an RMSE of 1.326.
The PCC exceeds those of most sequence-based, geometric, and pretrained
baselines included in Table~\ref{tab:lba_results_mimagent}, while remaining
slightly below BindNet. The RMSE is lower than the values reported for the
listed comparison methods. Because LBA30 imposes the more stringent
sequence-identity threshold, these results provide evidence that the selected
mathematical representation retains predictive information under reduced
protein homology between training and test sets.
On LBA60, MathAgent achieves a PCC of 0.787 and an RMSE of 1.262,
placing its performance numerically close to BindNet and SableBind and above
several earlier reference methods. 

Overall, the LBA results complement the CASF scoring-power evaluation by
showing that dataset-adaptive invariant selection remains effective under
sequence-identity-constrained protein--ligand affinity prediction.

\begin{table*}[htbp]
\centering
\caption{
Performance on the Atom3D-derived LBA30 and LBA60 sequence-identity
benchmarks \cite{townshend2020atom3d}. RMSE is reported in $pK_d$ units.
Reference methods are ordered by LBA30 PCC from highest to lowest; MathAgent
is shown separately for comparison.
}
\label{tab:lba_results_mimagent}
\small
\begin{tabular}{lcccc}
\toprule
\multirow{2}{*}{Method}
& \multicolumn{2}{c}{LBA30}
& \multicolumn{2}{c}{LBA60} \\
\cmidrule(lr){2-3}
\cmidrule(lr){4-5}
& PCC & RMSE & PCC & RMSE \\
\midrule

BindNet \cite{feng2023protein}
& 0.632 & 1.340 & 0.793 & 1.230 \\

SableBind \cite{li2025harnessing}
& 0.579 & 1.527 & 0.802 & 1.246 \\

GeoSSL \cite{liu2022molecular}
& 0.577 & 1.451 & -- & -- \\

Uni-Mol \cite{zhou2023uni}
& 0.565 & 1.434 & 0.753 & 1.357 \\

ProNet \cite{wang2022learning}
& 0.551 & 1.463 & 0.765 & 1.343 \\

Atom3D-CNN \cite{townshend2020atom3d}
& 0.550 & 1.416 & 0.608 & 1.621 \\

Atom3D-GNN \cite{townshend2020atom3d}
& 0.545 & 1.601 & 0.743 & 1.408 \\

HoloProt \cite{somnath2021multi}
& 0.509 & 1.464 & 0.749 & 1.365 \\

DeepDTA \cite{ozturk2018deepdta}
& 0.472 & 1.866 & 0.666 & 1.762 \\

SMT-DTA \cite{pei2022smt}
& 0.458 & 1.574 & 0.758 & 1.347 \\

ProtTrans \cite{elnaggar2021prottrans}
& 0.438 & 1.544 & 0.595 & 1.641 \\

DeepAffinity \cite{karimi2019deepaffinity}
& 0.415 & 1.893 & -- & -- \\

Atom3D-ENN \cite{townshend2020atom3d}
& 0.389 & 1.568 & 0.623 & 1.620 \\

TAPE \cite{rao2019evaluating}
& 0.338 & 1.890 & 0.568 & 1.633 \\

\midrule

MathAgent
& 0.623 & 1.326 & 0.787 & 1.262 \\

\bottomrule
\end{tabular}
\end{table*}

\subsection{Quantitative Toxicity Prediction}
\label{sec:toxicity_results}

\begin{table*}[htbp]
\centering
\caption{
Performance comparison on four quantitative toxicity benchmarks
\cite{martin2016user,wu2018quantitative,gao20202d}.
Squared Pearson correlation coefficient ($\mathrm{PCC}^{2}$) and RMSE are reported when available. MathAgent results are
obtained from a single evaluation of the fixed test set after the invariant
combination has been selected using training-side evidence. A dash indicates
that the corresponding metric was not reported in the source study.
}
\label{tab:toxicity_results_mimagent}
\small
\setlength{\tabcolsep}{3.5pt}
\begin{tabular}{lcccccccc}
\toprule
\multirow{2}{*}{Method}
& \multicolumn{2}{c}{IGC50}
& \multicolumn{2}{c}{LC50}
& \multicolumn{2}{c}{LC50DM}
& \multicolumn{2}{c}{LD50} \\
\cmidrule(lr){2-3}
\cmidrule(lr){4-5}
\cmidrule(lr){6-7}
\cmidrule(lr){8-9}
& $\mathrm{PCC}^{2}$ & RMSE
& $\mathrm{PCC}^{2}$ & RMSE
& $\mathrm{PCC}^{2}$ & RMSE
& $\mathrm{PCC}^{2}$ & RMSE \\
\midrule

Hybrid model \cite{moe2020development}
& 0.810 & --
& 0.678 & --
& 0.616 & --
& 0.629 & -- \\

3D MT-DNN consensus \cite{wu2018quantitative}
& 0.802 & 0.438
& 0.789 & 0.677
& 0.678 & 0.978
& 0.653 & 0.568 \\

2D MTDNN consensus \cite{gao20202d}
& 0.794 & 0.457
& 0.765 & 0.718
& 0.725 & 0.935
& 0.639 & 0.549 \\

BTAMDL \cite{jiang2020boosting}
& 0.793 & 0.447
& 0.778 & 0.694
& 0.741 & 0.934
& 0.638 & 0.580 \\

AweGNN \cite{szocinski2021awegnn}
& 0.789 & 0.451
& 0.737 & 0.753
& 0.541 & 1.171
& 0.667 & 0.557 \\

2D GBDT consensus \cite{gao20202d}
& 0.785 & 0.457
& 0.715 & 0.783
& 0.486 & 1.239
& 0.679 & 0.580 \\

T.E.S.T. consensus \cite{martin2016user}
& 0.764 & 0.475
& 0.728 & 0.768
& 0.739 & 0.911
& 0.626 & 0.594 \\

FDA \cite{martin2016user}
& 0.747 & 0.489
& 0.626 & 0.915
& 0.565 & 1.190
& 0.557 & 0.657 \\

Estate2 \cite{gao20202d}
& 0.742 & --
& 0.662 & --
& 0.623 & --
& 0.589 & -- \\

Estate1 \cite{gao20202d}
& 0.735 & --
& 0.694 & --
& 0.684 & --
& 0.605 & -- \\

Hierarchical \cite{martin2016user}
& 0.719 & 0.539
& 0.710 & 0.801
& 0.695 & 0.979
& 0.578 & 0.650 \\

Daylight \cite{gao20202d}
& 0.717 & --
& 0.724 & --
& 0.694 & --
& 0.617 & -- \\

Group contribution \cite{martin2016user}
& 0.682 & 0.575
& 0.686 & 0.810
& 0.671 & 0.803
& -- & -- \\

FP2 \cite{gao20202d}
& 0.681 & --
& 0.609 & --
& 0.357 & --
& 0.631 & -- \\

MACCS \cite{gao20202d}
& 0.643 & --
& 0.608 & --
& 0.434 & --
& 0.643 & -- \\

Nearest neighbor \cite{martin2016user}
& 0.600 & 0.638
& 0.667 & 0.876
& 0.733 & 0.975
& 0.557 & 0.656 \\

\midrule

MathAgent
& 0.733 & 0.508
& 0.724 & 0.772
& 0.471 & 1.267
& 0.629 & 0.590 \\

\bottomrule
\end{tabular}
\end{table*}

Quantitative toxicity prediction is a fundamental component of chemical
safety assessment and drug discovery, where the objective is to estimate the
adverse biological effects associated with chemical exposure. We evaluated
MathAgent on four established quantitative toxicity benchmarks: IGC50,
LC50, LC50DM, and LD50. 

MathAgent exhibits distinct predictive performance across the four toxicity
endpoints. On IGC50, the selected representation achieves
$\mathrm{PCC}^{2}$ of 0.733 and RMSE of $0.508$, placing its performance within
the range of several established molecular-descriptor and toxicity-prediction
approaches\cite{martin2016user,wu2018quantitative,gao20202d,
szocinski2021awegnn,moe2020development}. For LC50, MathAgent attains
$\mathrm{PCC}^{2}$ of 0.724 and RMSE of $0.772$. The corresponding correlation
matches that reported for the Daylight representation and is close to the
T.E.S.T. consensus, which yields $\mathrm{PCC}^{2}$ of 0.728 and
RMSE of $0.768$. For LD50, the largest toxicity dataset considered here, MathAgent obtains
$\mathrm{PCC}^{2}$ of 0.629 and RMSE of $0.590$. Its correlation is comparable
to those reported for FP2, the T.E.S.T. consensus, and the Hybrid model,
although several multitask, ensemble, and graph-based approaches achieve
higher values. These results indicate that the adaptively selected mathematical representations retain useful predictive signal across several toxicity endpoints, although their performance is more variable than on the protein--ligand benchmarks and does not uniformly match task-specific models.

LC50DM represents a distinct small-data regime, containing only 283 training
molecules and 70 test molecules \cite{wu2018quantitative}. Because further
subsampling would substantially reduce the information available for
representation screening, all 283 training compounds were used during
candidate evaluation. The leading invariant combinations achieved similar
training-side performance, providing limited evidence for a stable candidate
ranking. Following the predefined selection protocol, EIC+FPRC+PH was
retained and achieved a $\mathrm{PCC}^{2}$ of 0.471 on the held-out test set.
This result highlights the greater uncertainty of representation selection in
data-limited settings. A natural future extension would be to couple MathAgent
with closed-loop experimental design, allowing uncertain predictions or
representation choices to guide the acquisition of informative new training
data \cite{chhetri2026closed}.

\begin{figure*}[ht]
    \centering
    \includegraphics[width=\textwidth]{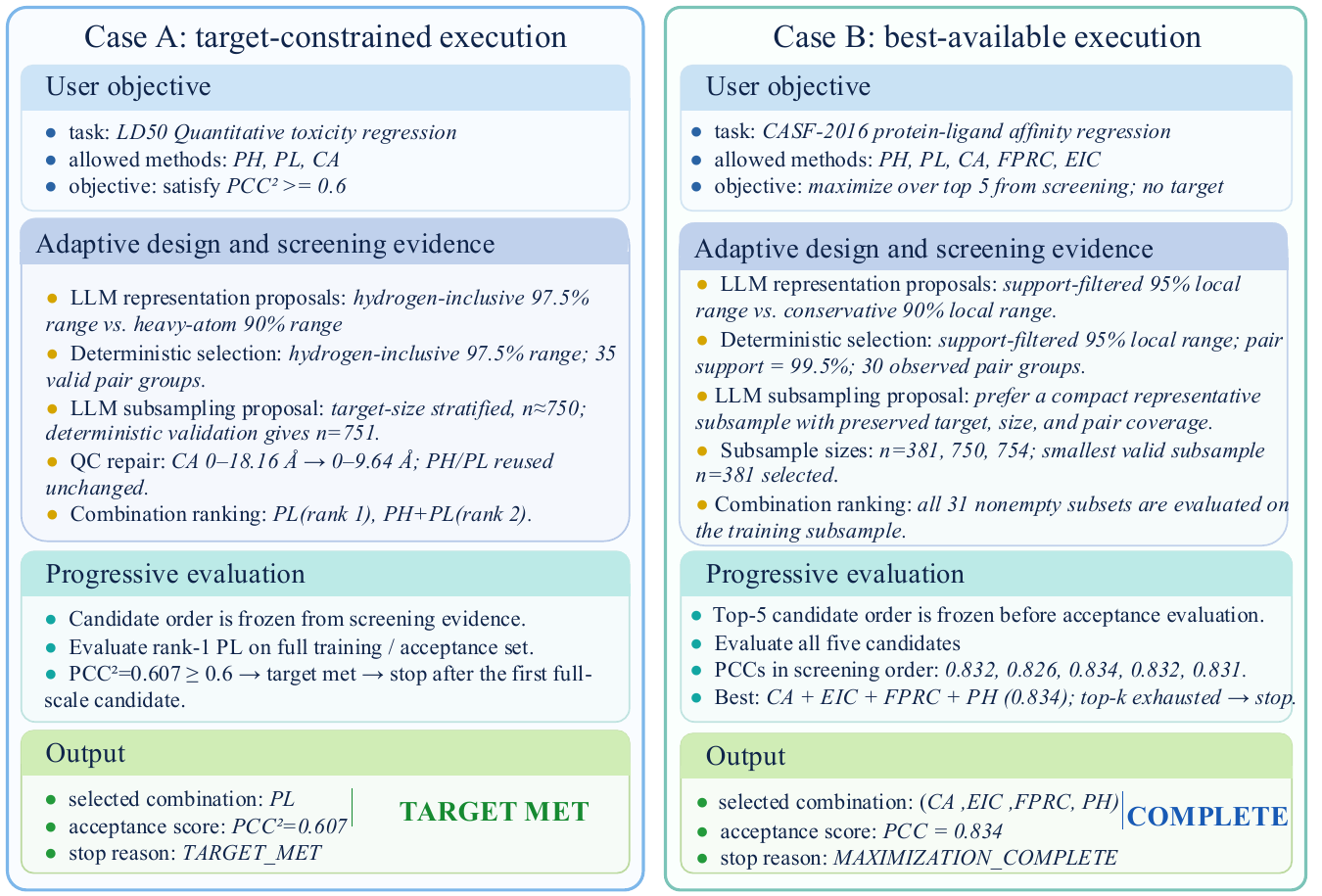}
   
\caption{
Objective-conditioned execution of MathAgent for target-constrained LD50 and
best-available CASF-2016 tasks, illustrating adaptive candidate selection and
stopping behavior under different user objectives.
}
    \label{fig:user_guided_modes}
\end{figure*}
\subsection{Objective-Conditioned Workflow Adaptation}
\label{sec:adaptive_execution}

In addition to predictive benchmarking, we evaluated whether MathAgent
adjusts its execution strategy in response to different user-specified
objectives. Figure~\ref{fig:user_guided_modes} contrasts two representative
settings: a target-constrained LD50 task and a best-available CASF-2016 task.
These analyses are intended to characterize objective-dependent workflow
behavior rather than to provide additional benchmark comparisons. Detailed
representation proposals, subsample construction, and quality-control decisions
for both executions are summarized in the figure.

For the target-constrained LD50 task, the admissible invariant families were
restricted to PH, PL, and CA, yielding seven nonempty candidate combinations,
and the user specified a target of
$\mathrm{PCC}^{2}\geq0.60$. After dataset-adaptive representation design and
training-side candidate screening, PL received the highest candidate priority,
followed by PH+PL. The candidate order was fixed before full-scale acceptance
evaluation. The first promoted candidate, PL, achieved
$\mathrm{PCC}^{2}$ of 0.607, satisfying the prescribed target. The workflow
therefore terminated after a single full-scale candidate evaluation rather
than continuing through the remaining lower-priority combinations. This case
demonstrates how a target-constrained objective can translate predictive
evidence into an explicit early-stopping decision and thereby limit
unnecessary full-scale computation.

The CASF-2016 task illustrates a different execution objective. All five
invariant families were admissible, producing 31 nonempty candidate
combinations, and the objective was to identify the strongest representation
within a predefined high-priority candidate pool rather than to satisfy a
fixed performance threshold. All 31 combinations were ranked using
training-only screening evidence, after which the top-five candidate pool was
frozen before acceptance evaluation. Each candidate in this fixed pool was
then evaluated, and CA+EIC+FPRC+PH achieved the highest acceptance PCC of
0.834. The workflow terminated only after the entire predefined pool had been
evaluated. In contrast to the target-constrained case, this execution allocates
additional computation to distinguishing among several high-priority
candidates rather than stopping when a sufficient performance level is first
reached.

The contrast between these two executions shows that MathAgent does not impose
a fixed evaluation trajectory across tasks. Instead, the same screening-based
prioritization framework supports different evaluation and stopping behaviors
according to the user-specified objective. The acceptance results shown here
illustrate adaptive workflow control and are distinct from protected-test
benchmark evaluation.

\subsection{Workflow-Level Computational Reduction}
\label{sec:efficiency}

We evaluated the computational efficiency of MathAgent using two workflow-level quantities: the amount of training data used for candidate screening and the number of candidate combinations promoted to full-scale evaluation. These quantities directly reflect the reductions achieved through training-set subsampling, candidate screening, and progressive candidate prioritization. Because end-to-end wall-clock time depends on hardware, system load, and the computational characteristics of individual invariant generators, we focus on these workflow-level measures rather than reporting a hardware-independent speedup. Representative results are summarized in Table~\ref{tab:efficiency}.

Training-set subsampling reduces the amount of data required for preliminary candidate screening when a representative subsample can be constructed. As shown in Table~\ref{tab:efficiency}, the LD50 and CASF--2016 executions used 751 of 5,931 and 381 of 3,772 training samples, respectively, for initial screening. The IGC50 execution used 753 of 1,434 samples, whereas the smaller LC50 dataset retained all 659 training samples. These results illustrate that subsample size is adapted to the available data rather than fixed across tasks, allowing screening cost to be reduced when subsampling is appropriate while avoiding unnecessary reduction for smaller datasets.

Screening-based prioritization further limits the number of candidate combinations requiring full-scale evaluation. With five admissible invariant families, the complete candidate space contains 31 nonempty combinations. Nevertheless, only three candidates were promoted to full-scale evaluation for IGC50, seven for LC50, and five for the CASF--2016 best-available execution. The reduction was more pronounced in the target-constrained LD50 case. Among seven admissible PH/PL/CA combinations, the highest-priority candidate, PL, achieved $\mathrm{PCC}^{2}=0.607$ and satisfied the user-specified criterion of $\mathrm{PCC}^{2}\geq0.60$, allowing the workflow to terminate after a single full-scale candidate evaluation.

In addition to reducing the amount of data used for screening and the number of full-scale candidate evaluations, MathAgent reuses previously generated invariant feature blocks whenever subsequent candidates require compatible representations. This avoids redundant feature generation across overlapping candidate combinations. Overall, the results in Table~\ref{tab:efficiency} show that MathAgent reduces computational workload by combining adaptive candidate screening with selective full-scale evaluation and representation reuse.

\begin{table}[t]
\centering
\caption{
Workflow-level computational reduction in representative MathAgent
executions. Subsample/train reports the number of samples used for training-side
screening relative to the full training set. Candidate counts report the
complete candidate space and the number subsequently evaluated at full scale.
}
\label{tab:efficiency}
\small
\setlength{\tabcolsep}{6pt}
\begin{tabular}{lcc}
\hline
Task / objective &
Subsample / train &
Candidates \\
&
&
space / full-scale \\
\hline
LD50, target-constrained
& 751 / 5931
& 7 / 1 \\

IGC50, best-available
& 753 / 1434
& 31 / 3 \\

LC50, best-available
& 659 / 659
& 31 / 7 \\

CASF-2016, best-available
& 381 / 3772
& 31 / 5 \\
\hline
\end{tabular}
\end{table}
\section{Concluding Remarks}
\label{sec:conclusion}

We developed MathAgent as an agentic framework for adaptive mathematical
representation selection in molecular property prediction. In the current
implementation, GPT-5.6 Terra supports task interpretation, dataset-aware
strategy proposal, and workflow coordination, while numerical scientific
operations are carried out by deterministic computational tools. This
separation enables flexible workflow adaptation while keeping feature
generation, model fitting, candidate ranking, and performance evaluation under
deterministic computational control.

Across the binding-affinity and toxicity benchmarks, MathAgent selected
different invariant combinations across datasets and prediction tasks. Within
protein--ligand affinity prediction, the preferred representations varied
between the conventional CASF scoring-power benchmarks and the
sequence-identity-constrained LBA benchmarks, while the toxicity endpoints
favored different invariant combinations. Objective-conditioned
executions further demonstrated that the same workflow can operate under
distinct user-defined goals, supporting target-constrained evaluation with
criterion-based early stopping as well as best-available evaluation over a
predefined candidate pool. These findings support a dataset-adaptive and
objective-aware approach to mathematical representation selection rather than
a fixed invariant composition or evaluation strategy.

An additional pattern emerges from the comparison between task types.
Relative to the corresponding reference methods, MathAgent is more competitive
on the protein--ligand binding-affinity benchmarks than on the small-molecule
toxicity benchmarks, where performance varies more substantially across
endpoints. One possible explanation is that the mathematical invariants
considered here are particularly well suited to structurally complex 
protein--ligand systems, where multiscale topology, geometry, spectral
structure, and element-resolved intermolecular organization provide multiple
sources of discriminative information. Individual small molecules are relatively structurally simple and may offer less pronounced multiscale structural organization for these sophisticated representations. Additionally, toxicity may depend more strongly on chemical information that is not explicitly encoded by the current invariant library. This interpretation should be regarded as a hypothesis rather than a mechanistic conclusion,
because the present experiments were not designed to isolate structural
complexity as a causal factor, and all experiments used the same fixed GBT
learner.

The study also highlights limitations of the current representation-selection
strategy. Training-side candidate prioritization becomes less reliable when
the available data are limited or when several representations achieve similar
predictive performance, as observed for LC50DM. Under such conditions, the
available evidence may be insufficient to establish a stable ordering among
near-tied candidates. Candidate rankings should therefore be interpreted
primarily as a mechanism for prioritizing subsequent computation rather than
as precise estimates of the relative superiority of closely matched
representations.

The principal contribution of MathAgent is a structured framework that
combines dataset-aware strategy generation, deterministic validation,
empirical candidate prioritization, computational reuse, and
objective-dependent execution within a unified representation-selection
workflow. Future work will extend the representation library beyond the
current mathematical invariants to include complementary physicochemical
descriptors, molecular fingerprints, and learned representations, allowing
MathAgent to determine when mathematical invariants are sufficient and when
hybrid representations are more appropriate. Future directions include
improving candidate selection in data-limited settings, incorporating
uncertainty when leading candidates are weakly separated,   extending the
current fixed-GBT setting toward architecture-aware invariant integration, and considering a wide range of scientific problems both within and beyond molecular sciences.

% \section*{Data and Code Availability}
%  All datasets supporting this study are publicly available. Protein--ligand structures and binding affinities were obtained from PDBbind (\url{https://www.pdbbind-plus.org.cn/}). 
%  %Docking decoy poses and GOLD scores are accessible at \url{https://weilab.math.msu.edu/AGL-Score}. The LBA30 and LBA60 datasets used for sequence-based generalization are derived from the Atom3D benchmark (\url{https://www.atom3d.ai/}).
%  Quantative toxicity datasets can be found in  \href{https://drive.google.com/drive/folders/1TfJOxHBlwttO6ve5wDkUSKplZXssS_2p?role=writer}{https://drive.google.com/drive/folders/1TfJOxHBlwttO6ve5wDkUSKplZXssS_2p?role=writer}.
% The code in this study can be found in  \href{https://github.com/yren24/MathAgent/tree/main}{https://github.com/yren24/MathAgent}.

\section*{Data and Code Availability}

All datasets used in this study are publicly available. Protein--ligand complex structures and experimentally measured binding affinities were obtained from the PDBbind database (\url{https://www.pdbbind-plus.org.cn/}). The quantitative toxicity datasets used in this study are publicly available at
\href{https://drive.google.com/drive/folders/1Nl9NgulYc08WjNnJcq1BCAmutCQH2TBb}
{https://drive.google.com/drive/folders/1Nl9NgulYc08WjNnJcq1BCAmutCQH2TBb}.

The source code and implementation used to reproduce the computational experiments in this study are publicly available at \href{https://github.com/yren24/MathAgent/tree/main}{https://github.com/yren24/MathAgent}.

 \section*{Acknowledgments}
 This work was supported in part by NIH grant  R35GM148196, the University of Georgia,   the Georgia Research Alliance, and the Defense Advanced Research Projects Agency (DARPA) under Agreement No. HR00112969E087.
The views and conclusions contained in this document are those of the authors, and should not be interpreted as representing the official policies, either expressed or implied, of the U.S. Government.

 \section*{AI Usage Declaration}

 AI was used to assist with sentence structure, grammar, and selected coding tasks during manuscript preparation and implementation. All scientific content, computational results, and final text were reviewed and approved by the authors. GPT-5.6 Terra was also used as the language-model component of the MathAgent framework to support task interpretation, dataset-aware strategy proposal, and workflow coordination, while numerical scientific computations were performed by deterministic tools.

\bibliographystyle{unsrtnat}
\bibliography{reference}
\end{document}